# Investigation of In Vitro Apocarotenoid Expression in Perianth of Saffron (*Crocus sativus* L.) Under Different Soil EC

*Investigasi Ekspresi Apocarotenoid secara In Vitro Bunga Saffron (Crocus sativus L.) pada EC Tanah yang Berbeda*

*Author(s):* Mandana Mirbakhsh[1]*; Zahra Zahed[2]; Sepideh Mashayekhi[2]; Monire Jafari[3]

[1] Department of Agronomy, Purdue University
[2] Department of Plant Physiology, Alzahra University
[3] Department of Mathematics, Texas State University,
*Corresponding author: *mmirbakh@purdue.edu*

## ABSTRACT

*Crocus sativus is a triploid sterile plant with red stigmas belonging to family of Iridaceae, and sub-family Crocoideae. Crocin, picrocrocin, and safranal are three major carotenoid derivatives that are responsible for color, taste and specific aroma of Crocus. Saffron flowers are harvested manually and used as spice, dye or medicinal applications. The natural propagation rate of most geophytes including saffron is relatively low. An in vitro multiplication technique like micropropagation has been used for the propagation of saffron. To understand the efficiency of this alternative and study the molecular basis of apocarotenoid biosynthesis/accumulation, the RT-PCR method was performed on perianth explants that were cultured on MS medium to observe the level of expression of zeaxanthin cleavage dioxygenase (CsZCD) gene during stigma development, and also the impact of soil EC on its expression. The present study was conducted at Plant molecular and physiology Lab, Alzahra University, Tehran, Iran during 2011-2013. Stigma-like structures (SLSs) on calli were collected from immature perianth explants from floral buds of corms that were collected from Ghaen city, and compared to (Torbat-e Haidariye, Mardabad and Shahroud cities) for investigating the impact of different soil EC on CsZCD expression. The results indicated that CsZCD gene was highly expressed in fully developed red SLSs in perianth of cultured samples of Shahroud with the highest salinity. In this research, a close relationship between soil EC and second metabolites regulation is studied. Overall, these results will pave the way for understanding the molecular basis of apocarotenoid biosynthesis and other aspects of stigma development in C. sativus.*

## ABSTRAK

*Crocus sativus* adalah tanaman mandul triploid dengan stigma merah dari keluarga Iridaceae, dan sub keluarga Crocoideae. Crocin, picrocrocin, dan safranal adalah tiga turunan karotenoid utama yang bertanggung jawab atas warna, rasa, dan aroma spesifik Crocus. Bunga saffron dipanen secara manual dan digunakan sebagai rempah-rempah, pewarna atau aplikasi obat. Tingkat perbanyakan alami sebagian besar geofit termasuk saffron relatif rendah. Teknik perbanyakan in vitro seperti mikropropagasi telah digunakan untuk perbanyakan saffron. Untuk memahami efisiensi alternatif ini dan mempelajari dasar molekuler dari biosintesis/akumulasi apocarotenoid, metode RT-PCR dilakukan pada eksplan perianth yang dikultur pada media MS untuk mengamati tingkat ekspresi gen zeaxanthin cleavage dioxygenase (CsZCD) selama perkembangan stigma dan juga dampak EC tanah pada ekspresinya. Penelitian ini dilakukan di Laboratorium Molekuler dan Fisiologi Tumbuhan, Universitas Alzahra, Teheran, Iran selama 2011-2013. Struktur seperti stigma (SLS) pada kalus dikumpulkan dari eksplan perianth yang belum matang dari kuncup bunga yang dikumpulkan dari kota Ghaen, dan dibandingkan dengan (kota Torbat-e Haidariye, Mardabad dan Shahroud) untuk menyelidiki dampak EC tanah yang berbeda pada ekspresi CsZCD. Hasil menunjukkan bahwa gen CsZCD sangat diekspresikan dalam SLS merah yang berkembang penuh di perianth sampel kultur Shahroud dengan salinitas tertinggi. Dalam penelitian ini, dipelajari hubungan erat antara EC tanah dan regulasi metabolit sekunder. Secara keseluruhan, hasil ini akan memberikan pemahaman dasar molekuler biosintesis apocarotenoid dan aspek lain dari perkembangan stigma di C. *Sativus.*

## INTRODUCTION

*Crocus sativus* (Saffron) is one of the most expensive spices in the world and is characterized by red stigmas that are dried for use in food seasoning, coloring, and medicine. Saffron cannot produce seeds and is propagated by corms slowly. Corms only survive for one season, produce cormlets and give rise to new plants (Deo, 2003). The high yield of saffron depends on multiple factors such as environmental conditions and the ability of plants to adjust their metabolism to tolerate environmental functions. Enough watering after flowering in late autumn to early spring, especially in rainfed conditions increases saffron’s production (Mirbakhsh & Hosseinzadeh, 2013). However, since conventional propagation methods are very slow, tissue culturing and micropropagation are great alternatives for the large-scale multiplication of saffron (Ascough et al., 2009).

Plant secondary metabolites are considered crucial for plant responses against stress and adaptation to the environment. Moreover, the greater activity of antioxidant enzymes has a key role in enhancing tolerance and protecting plants against oxidative reactions (Mirbakhsh & Sohrabi, 2022). During saffron life stages, the color of the stigma passes through yellow and orange to red with the conversion of amyloplasts to chromoplasts and secondary metabolites biogenesis (Bouvier et al., 2003). The significant quantities of carotenoid derivatives, which are formed from the oxidative cleavage of β -carotene and zeaxanthin, accumulated in scarlet stigmas are used to express the quality of saffron (Cazzonelli, 2011; Moraga et al., 2004, 2009). Carotenoids, which have been extensively studied in stigma tissues are terpenoids and can be synthesized through two different pathways: (1) Mevalonic acid (MVA) occurs in the cytoplasm (Castillo et al., 2005) and (2) Non-mevalonic (MEP) pathway takes place in plastids (Arigoni et al., 1997; Rohmer et al., 1993).

Mevalonate synthesis initiates with three acetyl CoA molecules and then continues with isopentenyl diphosphate (IPP) molecules production, geranyl geranyl pyrophosphate (GGPP) (Naik et al., 2003), colorless phytoene, colored lycopene, β -carotene, (Britton et al., 1998) and zeaxanthin (Bouvier et al., 2003) in MVA pathway. Many enzymes which are catalyzing the reactions and are coded by related key genes such as Phytoene synthase (PSY), lycopene-ß-cyclase (LYC), β -carotenoid hydroxylase (BCH), and 7, 8 (7', 8') by zeaxanthin cleavage dioxygenase (ZCD). β-carotene with two rings is built up via the cyclization of lycopene with lycopene-β-cyclase (LYC) (Britton et al., 1998). The hydroxylation of β-carotene in the MVA pathway is catalyzed by β-carotenoid hydroxylase coded by the BCH gene to yield zeaxanthin (Castillo et al., 2005). The biogenesis of the color and odor active compounds of saffron is derived by bio-oxidative cleavage of zeaxanthin (Moraga et al., 2009) at points 7, 8 (7', 8') by zeaxanthin cleavage dioxygenase (CsZCD) (Pfander & Schurtenberger, 1982) to produce crocetin dialdehyde and picrocrocin (Figure1).

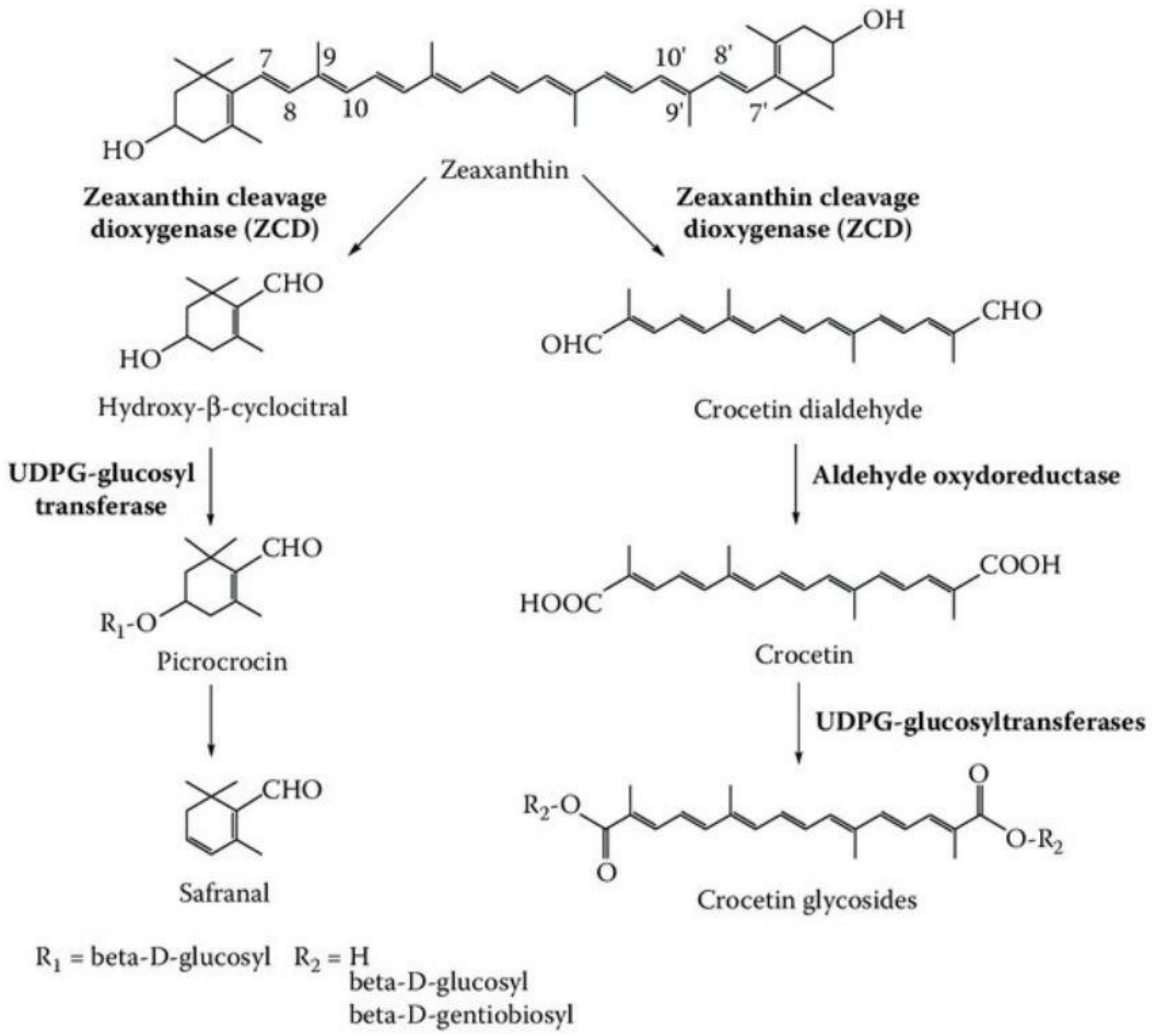

Figure 1. The biogenesis of the crocin and picrocrocin are derived from the bio-oxidative cleavage of Zeaxanthin at points 7, 8-(7', 8') (Naik et al., 2003).

In *C. sativus* stigmas, the final step involves glucosylation of the generated zeaxanthin cleavage products by glucosyltransferase 2 enzyme which is coded by the CsUGT2 gene in chromoplast of stigmas and then sequestered into the central vacuole of the fully developed stigma (Bouvier et al., 2003).

We aim to investigate the expression levels of the selected key gene, CsZCD, for apocarotenoid biosynthesis in *Crocus sativus* via the MVA pathway from *in-vitro*-formed stigma-like structures of the perianth. Moreover, we investigate the effect of different soil EC on CsZCD gene expression, which has a key role in the initiation of the synthesis of saffron pigment and aroma and is expressed specifically in the chromoplasts during the active period of zeaxanthin cleavage.

## METHODOLOGY

### Soil electrical conductivity (EC)

Soil samples (200g) from each of the four different mentioned regions were saturated with distilled water and mixed to consistent paste to determine the EC (Page et al., 1983). The sample dishes were sealed with parafilm and incubated at room temperature for one hour in order to diffuse water from the paste. The electrical conductivity of diffused water from the paste was measured by an electrical conductivity meter MODEL FE20–FiveEasy™ (Bremner, 2018).

The electric conductivity of the soils was analyzed and compared according to the different salinity classes that are shown in table 1 (USDA, 2014).

Table 1. Different classes of salinity according to EC (1 dS/m = 1 mmhos/cm) (USDA, 2014)

| Salinity class | EC (dS/m) |
|---|---|
| Non-saline | 0<2 |
| Very slightly saline | 2<4 |
| Slightly saline | 4<8 |
| Moderately saline | 8<16 |
| Strongly saline | 16≥ |

**Plant materials**

Saffron corms with attached immature flora buds were collected from the saffron field of Ghaen from November to December 2011. Flora buds were prepared for tissue culturing and compared with samples of other three distinct regions (Torbat-e Haidariye, Mardabad, and Shahroud) that were collected, prepared, cultured, and kept at -20 °C since 2010 (Mashayekhi & Hosseinzadeh Namin, 2015).

**Sterilization and culture**

Separated immature flora buds were thoroughly washed in running tap water (30 min) and then sterilized. Briefly, the buds were sterilized with 0.5% benzalkonium chloride solution for 15 min, treated first with 70% ethanol for 2 min, and followed with 5% sodium hypochlorite solution with a few drops of Tween 80 for 20 min (Namin et al., 2010). The buds were washed three times with sterile distilled water. For culturing, perianth organs were separated from sterilized buds and then cultured on MS medium with NAA (10mg/L), BAP (10mg/L), and sucrose 30% (Murashige & Skoog, 1962). All cultures were kept under 22 ± 2°C temperatures in darkness. Samples were sub-cultured every 28 days and collected from three development stages, freeze-dried, and then stored at -80°C for future analysis. Fully developed SLSs were chosen from four regions with different soil EC to be compared and analyzed.

**SLSs production measurement**

The number of SLSs was measured during a period of 3-6 months since tissue culture was initiated. Then calli with SLSs were collected, frozen in liquid nitrogen, and kept at -80°C for further analysis. The images of samples were taken by G6 Cannon digital camera from selected samples.

**RNA extraction and cDNA preparation**

Frozen stigmas from dried calli of four regions were ground in a cold and sterilized mortar and pestle into a fine powder and total RNA was extracted using an RNeasy Plant Mini kit (Qiagen, USA) following the manufacturer's protocol (Mirbakhsh & Sohrabi, 2022). The quality of the extracted RNAs was checked by measuring the absorbance at 260 and 280 nm by a spectrophotometer (Milton Roy-Spectronic 601-Co USA) and RNAs with a ratio of OD260/OD280 ranging from 1.2 to 1.5 were used for cDNA synthesis and ribosomal RNA profile were visualized with ethidium bromide staining following agarose gel electrophoresis. For preparing cDNA, 5-10 µg of each RNA sample as a template and 18 bp oligo dT primer and first strand cDNA synthesis kit RTpreMix (Bioneer Korea) were used. The synthesized cDNA was stored at -20°C for gene expression study.

**RT- PCR**

Reverse transcription was carried out for amplification of the CsZCD gene and CsTUB as internal control, according to the manufacturer's instructions. Gene-specific primers were designed (Bouvier et al., 2003; Castillo et al., 2005) to flank introns.

Designed forward and reverse primers were 5'-GTCTTCCCCGACATCCAGA TC-3' and 5'-TCTCTATCGGGC TCACGTTGG-3' for CsZCD gene with the length of 241 (GenBank access No. AJ489276) and 5'-ATGATTTCCAACT CGACCAGTGTC3' and 5'-ATACTCAT CACCCTCGTCACCATC-3' for CsTUB gene as control with the length of 225 bp (GenBank access No. AJ489275). PCRs were performed (Techne-Touchgene Gradient- FTG RAD 2D-LTD-UK), briefly according to the following conditions: 2 - 5 µg of cDNA was used. Initial denaturing at 95°C for 5min followed by 35 cycles of amplification according to the subsequent scheme; denaturing 1 min at 94°C, annealing at 56.2°C for 30 s and extension at 72°C for 40 s and final extension at 72°C for 10 min. The experiments were repeated twice. Subsequently, 7 l of the PCR products were used on 1% (w/v) agarose gel electrophoresis. The images of stained gels with ethidium bromide were scanned and captured by a gel documentation System.

**Statistical analyses**

All analyses were conducted with SPSS software (Version 15) and the means were compared by the Tukey method (95% confidence interval).

**RESULT AND DISCUSSIONS**

According to the results of soil EC measurements, Shahroud's soil was recognized as 'slightly saline' with EC=4.1, Torbat Heidariyeh with EC=2.2, and Ghaen with EC= 3.1 categorized as 'very slightly saline', and Mardabad with EC= 0.8 identified as a 'non-saline' soil. Exchangeable sodium ions at high levels lead to diffusion into the soil colloidal particles, so it destroys soil structure, breaks up the soil pores, and limits the soil drainage ability. Generally, neutral to slightly alkaline soil with electrical conductivity between 0.009 and 0.30 $dsm^{-1}$ is suitable for the growth and cultivation of saffron (Namin et al., 2010). Shahroud imposed very slight salt stress on the corms and Mardabad was the most suitable region with the lowest EC among the others. It seems that soils from Sharhrood (high sodium content and high EC) had more influence on corms properties as a main organ which is directly challenged with soil stress.

The first sign of callus induction on perianth explants was identified by inflation and transformation of explants during the period of 1 to 2 months (Fig. 2a). The induced calli went through three stages during their development from colorless calli without SLSs (stage I) (Fig. 2 b), to calli with pale yellow SLSs (Stage II) during the period of 2 to 4 months (Fig. 2 c), and fully developed scarlet SLSs during the period of 4 to 6 months when the explants were cultured (Stage III) (Fig. 2 d). These stages were comparable with those obtained from in vivo studies by Himeno & Sano (1987). They considered three developmental stages in *C. sativus* stigmas based on length, pigmentation, and apocarotenoid content.

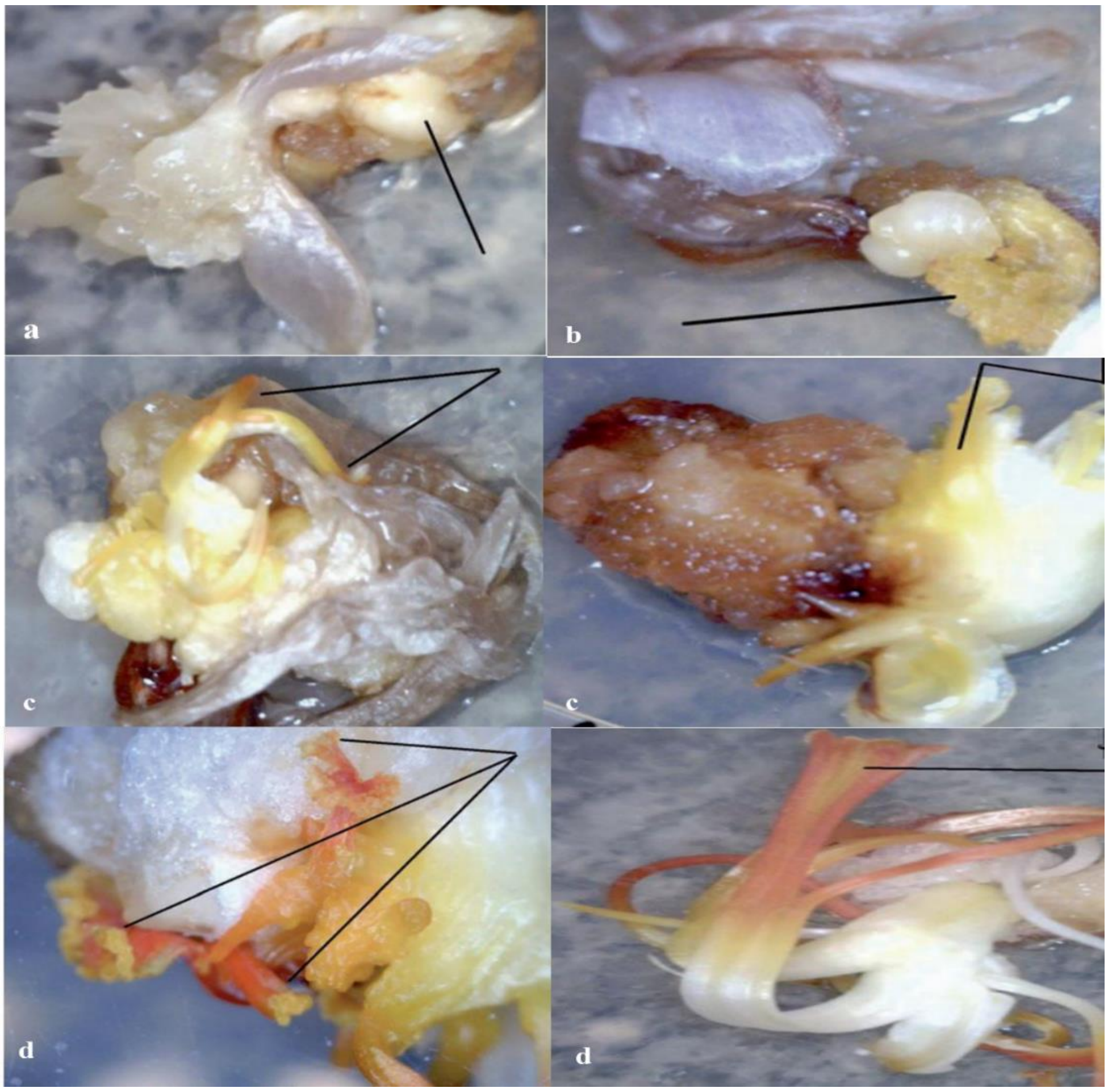

Figure 2. (a). The swollen cut edge of style, (b) Yellow stigma-like structures (SLSs) on callus of perianth with leaf-like structures, (c) Incomplete flower bud, and (d) Trumpet red stigma-like structures (SLS) on calli of style (Rs).

The SLSs percentages were 54, 40, 43, and 35 % for Mardabad, Torbat-e Heidariye, Ghaen, and Shahroud at the end of the given period (6 months) (Fig 3). The number of SLSs production was statically higher in Mardabad (non-saline) in comparison to the other three regions at the level of 0.05. The results obtained at the end of the experiment showed that Mardabad's had the highest SLSs percentage in comparison with Torbat-e Heidariye, Ghaen, and Sharood samples. This is in line with our results and signifies the adverse effect of salt stress on the quality of saffron stigmas. Moreover, this stress may influence metabolite reservations of corms in many ways which Effect of soil EC on saffron in vitro yield include epigenetic modification, and subsequently affect the quality and quantity reservoir of style which subsequently, affect the production of calli and SLSs when these style explants are used for tissue culture.

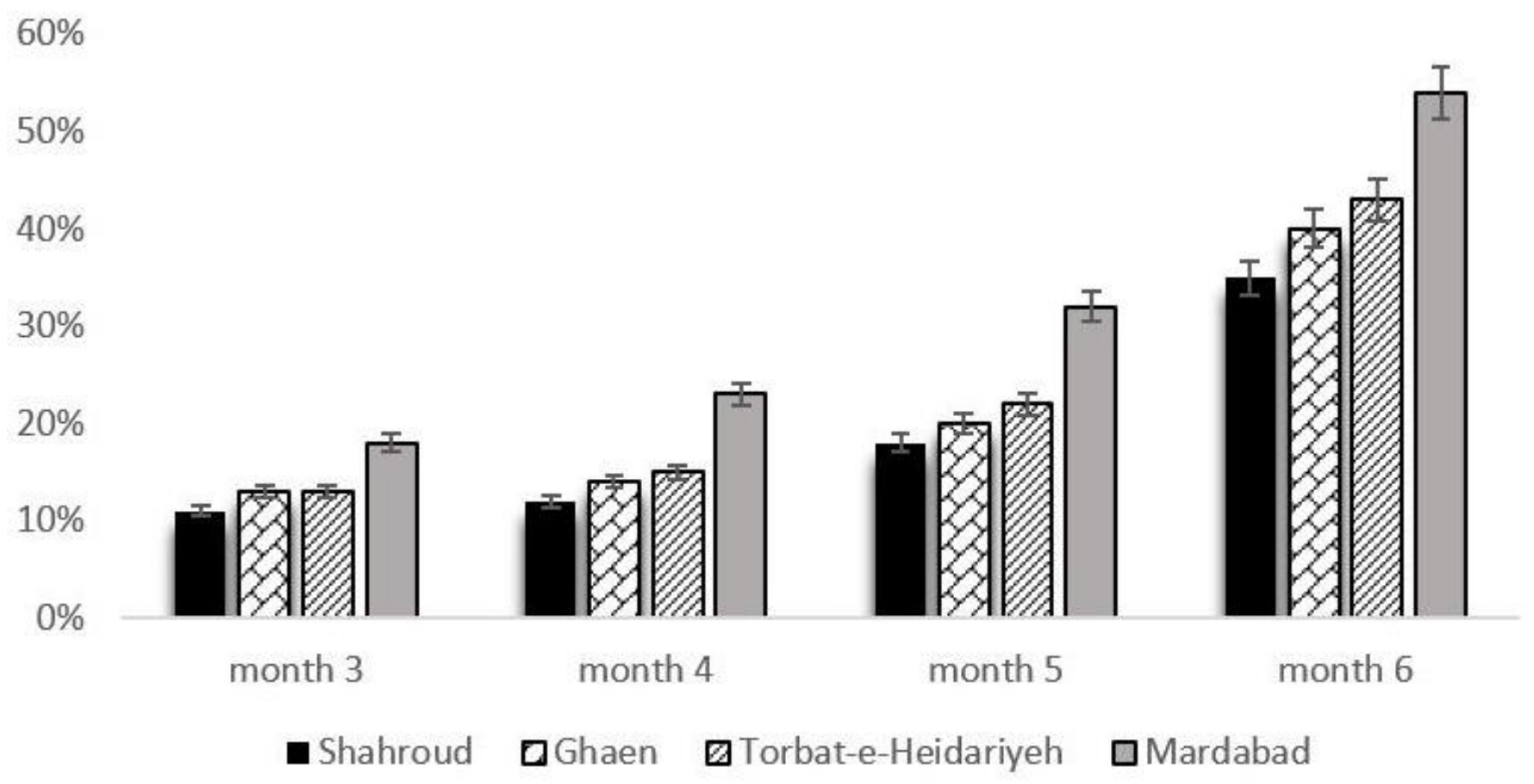

Figure 3. Percentage of stigma-like structures (SLSs) in Shahroud, Ghaen, Mardabad, and Torbat-e Heidariye, during 6 months.

Our results showed different expression levels of CsZCD between the regions with different soil ECs (Figure 4).

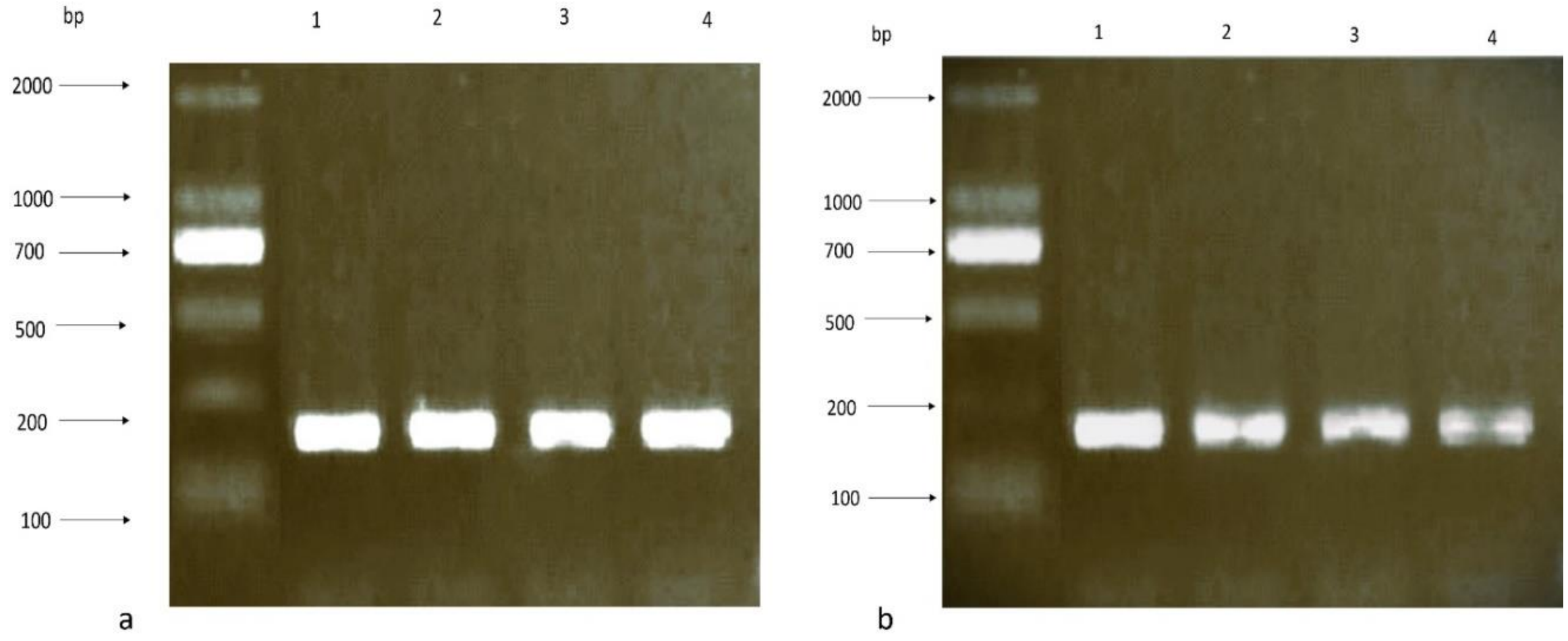

Figure 4. (a) Expression levels of Tubulin gene (CsTUB) in *C. sativus* as an internal control at four regions. (b) Different levels of zeaxanthin cleavage dioxygenase (CsZCD) expression with 241 bp at four regions (1) Shahroud, (2) Ghaen, (3) Torbat-e-Heydariyeh, (4) Mardabad.

The carotenoid pathway gene expression is affected by abiotic stressors such as light, water availability, salinity, and temperature (Moshtaghi et al., 2010). CsZCD has a regulatory role in apocarotenoid biosynthesis, content, and stigma development in saffron (Mir et al., 2012a, 2012b) and is highly expressed in the scarlet stigma with the highest level of zeaxanthin and apocarotenoids (Castillo et al., 2005). Our results showed a different level of expression among regions. The weakest expression was identified in Mardabad (non-saline), while CsZCD was upregulated as soil EC increased in Shahroud (slightly salinity), which raise the protective role of the phenolic compound under abiotic stress (Waśkiewicz et al., 2013) (Figure 4b). However, the decrease in percentage SLSs production under a slight saline region (Shahroud) indicated the adverse effect of salt stress on the quality of saffron stigmas, which might therefore be due to the redirection of apocarotenoid procures to hormone abscisic acid (ABA) production (Tuan et al., 2013). The production of plant hormone ABA, as a derivative of apocarotenoids, is increased in response to environmental stresses such as drought, salinity, and high/low temperature (Tuteja, 2007). This finding suggested that upregulation of CsZCD improves the ability of a plant to alleviate oxidative stresses under salinity conditions.